\documentstyle[11pt,newpasp,twoside,epsf]{article}
\def\uu{\raisebox{-.1cm}{$_\sqcup$}}
\def\sla{\mbox{$\stackrel{<}{{\mbox{\tiny $\sim$}}}$}}

\def\edcomment#1{\iffalse\marginpar{\raggedright\sl#1\/}\else\relax\fi}
\reversemarginpar

\begin{document}

\title{Handling of Atomic Data}
\author{Thomas Rauch}
\affil{Dr.\,Remeis-Sternwarte Bamberg, Sternwartstra\ss e 7, 96049 Bamberg, Germany; Institut f\"ur Astronomie und Astrophysik, Sand 1, 72076 T\"ubingen, Germany}
\author{Jochen L\@. Deetjen}
\affil{Institut f\"ur Astronomie und Astrophysik, Sand 1, 72076 T\"ubingen, Germany}

\begin{abstract}
State-of-the-art NLTE model-atmosphere codes have arrived at a high level of
numerical sophistication and are now useful tools to analyze high-quality
spectra from the infrared to the X-ray wavelength range. The capacity of
current computers permit calculations which include line spectra from all
elements from hydrogen up to the iron group. The lack of reliable atomic data
has become a critical problem for further progress.

We summarize available sources of atomic data, and discuss how these are 
implemented in the T\"ubingen Model-Atmosphere Package \linebreak \textsc{tmap}. We
describe our Iron Opacity Interface {\sc IrOnIc} which is used to calculate
opacities of iron-group elements from Kurucz's and the Opacity Project's data.

We propose general use of the T\"ubingen Model-Atom Database \linebreak {\sc tmad}, which
would allow an easy exchange of ready-to-use model atoms between all
model-atmosphere groups. The comparison of model-atmo\-sphere calculations would
then be much easier, and would save a great deal of manpower that is presently
consumed preparing suitable model atoms for spectral analyses.
\end{abstract}

\section{Introduction}
In the early 80's of the last century, the implementation of approximate
lambda-operators (ALO, leading to ``accelerated lambda iteration'', ALI) 
in the NLTE model-atmosphere codes at Kiel by Werner \& Husfeld (1985) and
Werner (1986) provided an efficient method to calculate synthetic stellar
spectra of hot stars. Together with the access to the fourth {\sc cray}
computer (a {\sc cray-1\,m}) in Germany (at the Konrad-Zuse-Zentrum f\"ur
Informationstechnik Berlin), which was installed in February 1984, powerful
tools for spectral analysis of hot star spectra were developed. At the end of
1987, the Rechenzentrum der Universit\"at Kiel installed a {\sc cray\,x-mp}.
Access to this machine (in the framework of the Norddeutscher 
Vektorrechner-Verbund) and the following, even more powerful, {\sc cray}
computers, made all our efforts possible.

The first of our NLTE codes, \textsc{hymoc} ({\sc HYd}rogen 
{\sc MO}del-atmosphere {\sc C}ode) (Werner 1986), was able to calculate only
pure-hydrogen model atmospheres. However, inasmuch as the atomic data for
hydrogen are well known, it was an ideal tool to investigate the numerical
approximations, and limitations to the size of model atoms, used in earlier
calculations, which were necessary before the ALI method was developed
(Rauch \& Werner 1988).

The next code \textsc{pro2} ({\sc pro}gram no.\,2) (Werner 1988; Werner \&
 Dreizler 1999) is much more flexible, and is able to take into account all
elements up to the iron group (Dreizler \& Werner 1993; Rauch 1997). 
It has been used successfully for the analysis of hot stars (e.g.~Rauch \&
Werner 1991; Rauch 1993; Rauch 2000). Our NLTE group moved from Bamberg (1993)
and Potsdam (1995) to T\"ubingen (since 1996) where \textsc{tmap}, the
state-of-the-art {\sc T}\"ubingen {\sc M}odel {\sc A}tmosphere {\sc P}ackage
was created. With the newly developed \textsc{IrOnIc} code (\S 4), it was
possible to calculate an extended grid of realistic stellar fluxes from models
which take into account the opacities from H -- Ni (Rauch \& Deetjen 2001).
These models are plane-parallel, in hydrostatic and radiative equilibrium, have
350 atomic levels which are treated in NLTE, about 1,000 individual lines from
H - Ca, and millions of lines from the iron-group elements.

The state of the field of spectral-analysis of hot stars has completely changed
within the last two decades. At the beginning of the 80's, the main obstacles
were insufficient numerical methods and computational capacities. Rauch \& 
Werner (1991) have shown the enormous progress which came with the ALI method,
with examples of very detailed H + He + C + N models in contrast to 
``classical'' H + He models. At present, the lack of reliable atomic data for
metals, line-broadening tables, etc.~set undesirable limits to highly-developed
NLTE codes. This lack often hampers an adequate analysis e.g.~of high-resolution
UV spectra provided by the STIS (Space Telescope Imaging Spectrograph) aboard
the Hubble Space Telescope (HST) or by the Far Ultraviolet Spectroscopic
Explorer (FUSE).

\section{Sources of Atomic Data}
For the elements H, He, C, N, and O, we use standard sources such as
Bashkin \& Stoner (1975, 1978) and Moore (1959, 1971), for energy levels, 
Wiese et al.~(1966, 1969) for transition probabilities, and input-data 
(mainly for C, N, and O) compiled for the program {\sc detail} (K.~Butler,
Munich, private communication). Most of the data can also be found e.g.
 \smallskip\\
in the National Institute of Standards and Technology (NIST) Atomic Spectra Database\\
\hbox{}\hspace{3mm}at {\tt http://physics.nist.gov/cgi-bin/AtData/main\_asd},\smallskip\\
in the Vienna Atomic Line Database (VALD) database\\
\hbox{}\hspace{3mm}at {\tt http://www.astro.uu.se/$^\sim$vald},\smallskip\\
in The Atomic Line List of Peter van Hoof\\
\hbox{}\hspace{3mm}at {\tt http://www.pa.uky.edu/$^\sim$peter/atomic/},\smallskip\\
or in the Kurucz Atomic Line Database\\
\hbox{}\hspace{3mm}at {\tt http://cfa-www.harvard.edu/amdata/ampdata/kurucz23/sekur.html.}\smallskip\\
We wish to call attention to the compilation of Wiese, Fuhr, \& Deters (1996)
who show the difficulties in judging the quality of atomic data for C, N, and O.

For the elements F -- Ca we use mainly data provided by the Opacity Project
(\S 2.1). For the iron-group elements, we use both the Kurucz's lists
(1993) and OP data.
A.~K.~Pradhan provides atomic data, and links to related data and some review
papers, e.g\@. Pradhan \& Peng (1995)

at {\tt http://www.astronomy.ohio-state.edu/$^\sim$pradhan}.

\subsection{Opacity Project Data}
In the framework of the OP (Seaton 1987), opacities for stellar envelopes were 
calculated during 1987 -- 1994. The OP work is documented in about 50
publications ({\tt http://heasarc.gsfc.nasa.gov/topbase/publi.html}), e.g.~20
alone in J. Phys. B. A ``final report'' was given by Seaton et al.~(1994). The
aim of the OP was to provide {\it access to a complete and accurate dataset},
consisting of energies, photoionization cross-sections, and oscillator
strengths.

{\sc TOPbase} ({\tt http://heasarc.gsfc.nasa.gov/topbase/topbase.html}), 
\linebreak the OP on-line atomic database, provides data for elements with 
$Z\in \{1 - 14, 16,\\
18, 20, 26\}$, and ions with $N_{\rm e} \in \{ 1 - 26\} $. Energy levels, hence
transitions are complete for orbital quantum numbers $l\in\{\rm{s, p, d, f}\}$. 
Because the level energies calculated by the OP are slightly different from
laboratory measurements (e.g.~Werner \& Rauch 1994), we replace them by Bashkin
\& Stoner (1975, 1978) values, if available.

A detailed investigation of OP photoionization cross-sections 
($\sigma^{\sc op}_{\rm bf}$) was made by Rauch (1997). The impact of 
 $\sigma^{\sc op}_{\rm bf}$ is significant, and I recommend use of the OP data.
In our application, the OP data is interpolated by \textsc{pro2} to the 
frequency grid used. It is not necessary to use a very fine grid in the 
model-atmosphere calculation (see \S 5) in order to resolve the resonances in 
$\sigma^{\sc op}_{\rm bf}$ -- because of the energy-level uncertainty (see
above), these are not at the correct wavelengths anyway. However, they
contribute to the transition rates 
$R_{ij}\sim\int_0^\infty(J_\nu\sigma_{ij}/h\nu)d\nu$
and one has to be aware that they are then neglected there. In the event we
have replaced the threshold energy, $\sigma^{\sc op}_{\rm bf}$ is shifted
accordingly.

Because an extrapolation of OP bound-free data into the high-energy range may
result in negative cross-sections, \textsc{pro2} estimates the slope of the
cross-section and, if necessary, replaces it by an approximation based on
Seaton's formula (1962)
\begin{equation}
\sigma_\nu = \sigma_0\left(\frac{\nu_{\rm th}}{\nu}\right)^s
             \left[\alpha + (1-\alpha)\frac{\nu _{\rm th}}{\nu}\right]
\end{equation}
\noindent
using $\alpha$ = 1, and $s$ = 3. $\sigma _{\rm 0}$ is then estimated from the
last available OP data points. Setting negative cross-sections to zero might
result in an arbitrary emission bump in what possibly has become an
``opacity-poor'' high-energy range of the synthetic spectrum.

\subsection{Iron Project data}
The Iron Project (IP, Nahar, these proceedings) has the goal of computing, on a
large scale, electron-excitation cross-sections and rates of astrophysical and
technological importance, using the most reliable procedures currently available
({\tt http://www.usm.uni-muenchen.de/people/ip/iron-project.html}). The \linebreak IP data
will be available in {\sc TIPTOPbase} in the near future. The energy levels are
much more accurate than those of the OP data. Thus we plan to use these data as
soon as possible. 

\section{Hydrogen and Helium}
Because of the limits of the accuracy of the numerical methods and computers
presently available, the number of levels which can be treated simultaneously
in NLTE by \textsc{pro2} is limited ($\sla$ 350). Thus, for every ion some
(lower) NLTE and some (higher) LTE levels must be selected. For example, in the
case of hydrogen a ``classical'' cutoff at a total number of 16 levels was used
by Auer \& Mihalas (1969). Rauch \& Werner (1988) found that in the case of
H\,I, eight NLTE levels yield an accuracy better than 1\% in the emergent flux.
A more accurate approach which eliminates the necessity of this cutoff has been
presented by Hubeny et al.~(1994). They generalized the occupation-probability
formalism (HMF) of Hummer \& Mihalas (1988) to NLTE conditions. Using exact
partition functions, they have shown that the number of levels is not an
indicator of the accuracy of a model. We generally use the HMF for hydrogen and
helium (Werner 1996). For other species it is used in an approximate way as
well.

A crucial point in model-atmosphere calculation are the collisional rates
--- even in the case of H\,I and He\,II, where atomic data are known much better
compared to all other atoms/ions. For the collisional excitation rates of H\,I,
we follow Sampson \& Golden (1970, and references therein) and Mihalas (1972)
\begin{equation}
C_{ij} = 4\pi a_0^2\,\sqrt{\frac{8k}{\pi m_e}}\,n_e\,T^{1/2} 
       \left(\frac{E_H}{h\nu_{ij}}\right)^2 
        f_{ij}\,u_0\,[\,E_1(u_0)+0.148\,u_0 E_5(u_0)\,]\,\gamma
\end{equation}
Here $f_{ij}$ is the optical oscillator strength of the transition 
$i\rightarrow j$, $E_H$ is the ionization energy of hydrogen, 
$u_0 \equiv h\nu _{ij}/kT, \gamma \equiv \beta + 2(\alpha - \beta)/\Delta n$
where $\beta = 3 - 1.2/n_i,\quad\alpha = 1.8 - 0.4/n_i^2$ if $\Delta n > 1$,
and $\gamma$ = 1 otherwise.
Mihalas (1967) remarked ``Clearly the cross-sections we have chosen are very
approximate, and we eagerly await better values from theory and experiment.''
for this formula. A detailed comparison with more recent data (e.g.~Jones,
Madison, \& Srivastava 1991) remains to be done. The collisional excitation
rates of helium are calculated by formulae of Mihalas \& Stone (1968). The He I
formulae that are used by {\textsc pro2} are summarized in Werner, Rauch, 
\& Dreizler (2001). For He II we use 
\begin{equation}
C_{ij} = \pi a_0^2\,\sqrt{\frac{8k}{\pi m_e}}\,n_e\,T^{1/2}
         \left(\frac{E_H}{h\nu_{ij}}\right)^2\,f_{ij}\,u_0\,e\,
	 [\,\ln 2\,e^{-u_0} + E_1(u_0)\,]\,\gamma
\end{equation}
with $\gamma = \min(n_i, 1.1)\times\min[\,\Delta n, n_i-(n_i-1)/\Delta n\,]$
(Mihalas 1972).
Mihalas \& Stone (1968) stated that ``Naturally these expressions are only
first approximations, but certainly they are right to order of magnitude.''.
As in the case of hydrogen, Jones, Madison, \& Srivastava (1991) have published
new data.

To summarize, model-atmosphere calculations are lacking reliable collisional
data. While for hydrogen and helium at least approximate formulae are available,
the situation for most metals is worse. If the optical oscillator strength of
a transition is known, we use the van Regemorter formula (\S 4.4). If it is
unknown, we use a formula given by Butler ({\tt http://ccp7.dur.ac.uk/Docs/detail.ps}).
\begin{equation}
C_{ij} = \pi a_0^2\,\sqrt{\frac{8k}{\pi m_e}}\, n_e\,T^{1/2}\,(1+u_0)e^{-u_0}.
\end{equation}

\section{\sc IrOnIc}
The Iron Opacity Interface \textsc{IrOnIc} allows one to create line
cross-sections and model-atom files for iron-group elements, i.e.~Ca to Zn.
The aim of \textsc{IrOnIc} is to prepare radiative bound-bound (RBB),
radiative bound-free (RBF), collisional bound-bound (CBB), and collisional
bound-free (RBF) cross-sections (CS) as input for \textsc{pro2}.

A detailed consideration of all atomic line transitions, known from experiment
or theoretical calculations would be impossible. The large number of levels
would exceed both the computational power, and overwhelm the numerical methods
for model-atmosphere calculations, available at the moment. Blanketing by
millions of lines from the iron-group elements arising from transitions between
some $10^5$ levels can be attacked only with statistical methods.

\subsection{The Basic Concept}

The basic concept for achieving this goal is to combine all energy levels
of one (model) ion into typically 6 -- 20  energy bands (Haas 1997, Fig.~1).
Whereby one can distinguish between levels with even and with odd parity.
This reduces the number of levels dramatically, without losing too much
information about the physics of the system (Anderson 1989; Anderson \&
Grigsby 1991). Each of these bands is then treated as a single NLTE level
with a suitably averaged energy $E_L$ and statistical weight $G_L$, which are
computed from the individual levels within a particular band (Eq.~5--7). 
An atom whose level structure is simplified in this way is called a model atom.
\begin{eqnarray}
  E_L &=& \sum_{l\epsilon L}\,E_{\,l}\,g^*_{\,l} 
                \left/\sum_{l\,\epsilon\, L} g^*_{\,l} \right. ,\\  
  G_L &=& e^{E_L/kT^*}\times\sum_{l\,\epsilon\, L} g^*_{\,l},\\
  g^*_{\,l} &=& a_s\,g_{\,l}\,e^{-E_{\,l}/kT^*}.
\end{eqnarray}
$E_{\,l}$ and $g_{\,l}$ are the energy and the statistical weight of a real
atomic level $l$. $L$ indicates a particular superlevel. $T^*$ is a typical
temperature, pre-chosen and fixed, given by the Saha ionization equilibrium,
where the particular ionization stage dominates. The basic assumption here is
that all individual levels within one band have the same NLTE departure
coefficient.

Additionally, \textsc{IrOnIc} permits one to combine several chemical species 
into one generic model atom. A ``generic model atom'' means it appears to be
one atom with several bands, but contains the co-added cross-sections of all
species included in it. The contribution of the individual species to the total
result is determined by their abundances $a_s$ with respect to the dominant
atom (e.g.~Fe/Fe = 1 and Ni/Fe = 0.5 etc.).
Furthermore, \textsc{IrOnIc} allows one to generate a combined model atom. That
means that one combines several model atoms for different elements. All
calculations must be done on one common frequency grid.

\begin{figure}[ht]
\plotone{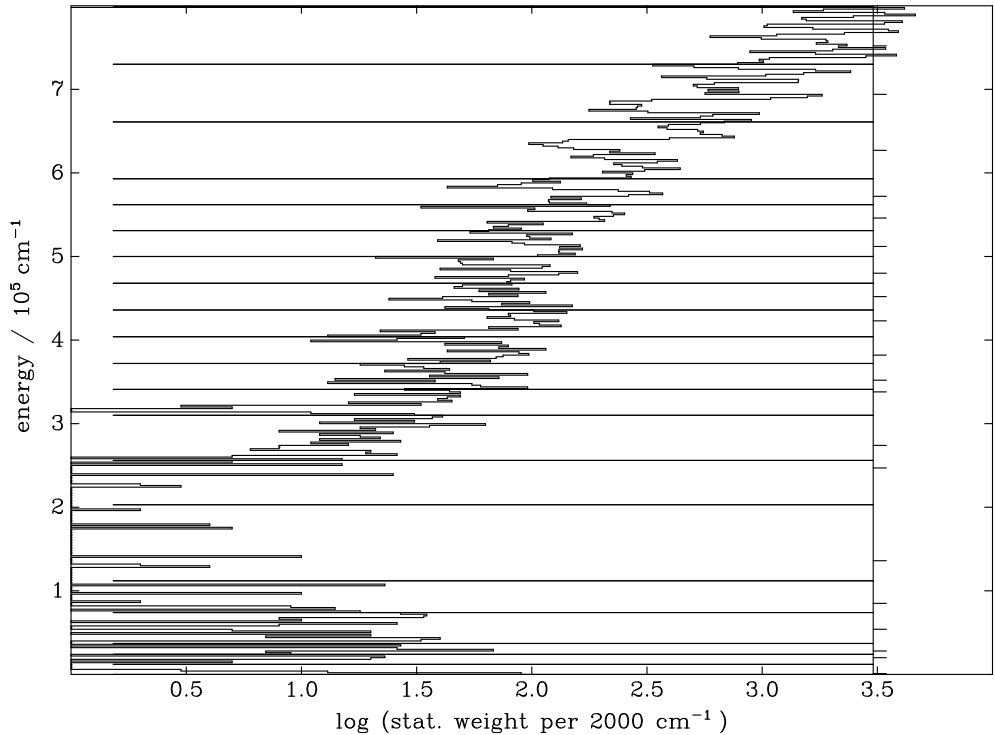}
\caption{Energy distribution of statistical weights of the iron-group
elements in ionization stage {\sc vi}. Individual energy levels are
grouped here into 20 bands (interval limits are indicated by horizontal lines) 
and merged into superlevels with an average energy (indicated by small tick
marks at right hand side).}
\end{figure}

\begin{figure}[ht]
\plotone{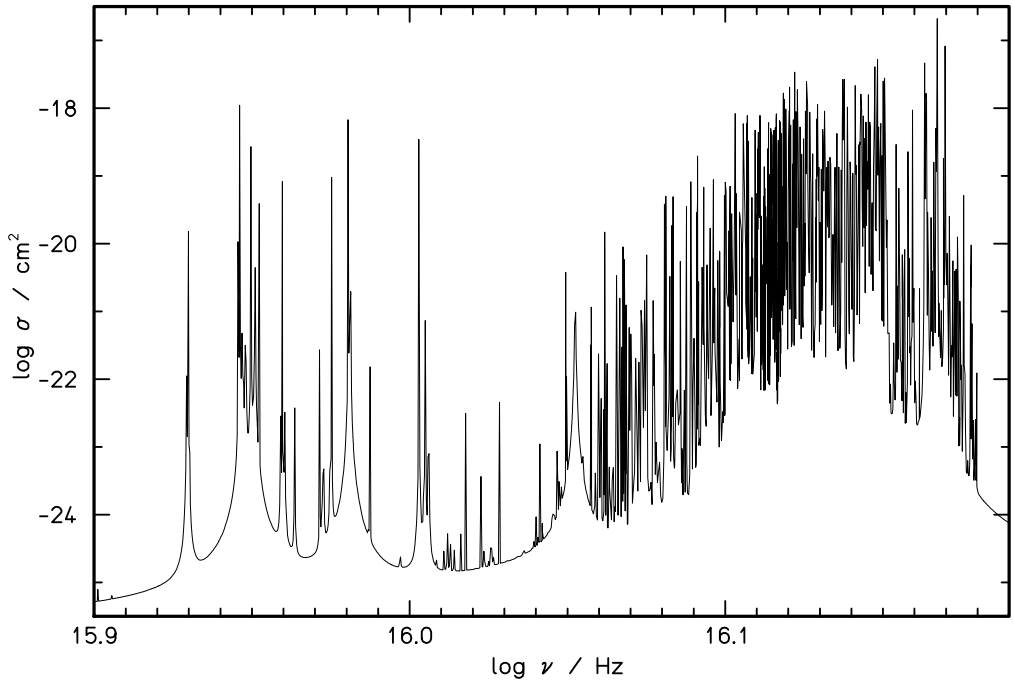}
\caption{Example of the band-band cross-section of ionization stage {\sc vi}
in a generic model atom which includes all iron-group elements. 2,289
individual lines between two superlevels are co-added to form a complex photon
cross-section which is sampled in 1,913 frequency points.
            }
\end{figure}

\subsection{The Atomic data}

Understanding the properties of model atoms and generic cross-sections requires
knowledge about the basic properties of the individual atoms. The calculation
of energy levels and transition probabilities by Kurucz (1993) and Seaton et
al.~(1994) are the most important sources of atomic data for our calculations.
Nevertheless \textsc{IrOnIc} has a flexible interface, which, in principle,
can handle all kinds of atomic data sources.

\subsection{Calculation of the RBB cross-sections}
The cross-section for a transition between two superlevels must be composed of
the individual cross-sections between levels in the upper and in the lower band.
This can't be done by co-adding all the individual oscillator-strengths
$f_{lu}$ into one ``super-oscillator-strength'', because this would result in
only one unrealistically strong ``superline''. Therefore the frequency-dependent
cross-sections of all individual lines are co-added in order to guarantee that
each individual line core is considered at the correct frequency position:
\begin{equation}
\sigma_{LU}=\frac{\pi e^2}{m_e c}\sum_{(l\,\epsilon\,L;\,u\,\epsilon\,U)}
g^*_{\,l}f_{\,lu}\,\phi(\nu_{\,lu}-\nu)\left/\sum_{l\,\epsilon\,L}g^*_{\,l}\right. .
\end{equation}
$\pi e^2/m_e c$ is the cross-section of a classical oscillator and $\phi$ is 
the normalized line-profile of the individual lines. The shape of the profile
function is a Voigt profile with its width determined by various broadening
mechanisms: natural line broadening, Doppler broadening, and Stark broadening.
For the iron-lines the quadratic Stark effect is of importance. It describes
the interaction between ions and electrons. According to Cowley (1971), the
resulting broadening of the individual lines is approximately given by:
\begin{equation}
\gamma_{Stark} = 5.5\times10^{-5}\;\frac{n_e}{\sqrt{T}}\;\left
    [\,\frac{(n_{\rm eff}^{\rm up})^2}{z+1}\,\right]^2.
\end{equation}
Where $n_{\rm eff}^{\rm up}$ is the effective principal quantum number of the
upper level, and $z$ is the effective charge seen by the active electron.
Because the iron-group cross-sections must be calculated in advance, the
electron density and the temperature within the atmosphere cannot be used here.
Therefore we calculate the broadening of the line, according to Rauch \& Werner
(1988), with a fixed temperature $T_{\,\rm line} = 3/4\times T_{\rm eff}$.  
The variation of the electron density within an atmosphere is taken into
account, by calculating $\gamma_{Stark}$ for two different electron densities:
namely 0 and $10^{16}$ electrons/cm$^3$. Because the Stark wings of the lines
depend linearly on electron density, the values of $\gamma_{Stark}$ for other
electron densities can be interpolated linearly by {\textsc pro2}.

In analogy to the inter-band cross-section $\sigma_{LU}$ an intra-band
cross-section $\sigma_{LL}$ can be defined:
\begin{equation}
\sigma_{LL} = \frac{\pi e^2}{m_e c}
\sum_{(l\,\epsilon\,L;\,l^{\prime}\,\epsilon\,L} g^*_l f_{l\,l^{\prime}}
\phi(\nu_{l\,l^{\prime}}-\nu)\left/\sum_{l\,\epsilon\,L}g^*_l\right.
\end{equation}
These cross-sections are used as additional background-opacities in the
radia\-tion-transport equation. This type of transition does not occur
when the individual energy bands include levels of the same parity only.

The resulting RBB transitions $\sigma_{LU}$ and $\sigma_{LL}$ show a complex,
frequency-dependent cross-section. An example of $\sigma_{LU}$ is shown in
Fig.~2. 

\subsection{The calculation of the CBB, RBF, and CBF cross-sections}

The CBB cross-sections are calculated with the formula of
van Regemorter (1962). For allowed dipole transitions we use

\begin{equation}
C_{ij} = \pi a_0^2\,\sqrt{\frac{8k}{\pi m_e}}\,n_e\,\sqrt{T}
         \left[\,14.5 f_{ij} \left(\frac{E_H}{E_0}\right)^2\right]
         u_0 e^{-u_0}\Gamma(u_0)
\end{equation}
where $u_0\equiv h\nu_{ij} / kT, E_0=h\nu_{ij}, 
\Gamma(u_0)={\rm max}\left[\,\bar{g},\,0.276\,e^{u_0}\,E_1(u_0)\,\right]$,
and $\bar{g}=0.2$ for $n'\neq n$, otherwise $\bar{g}=0.7$.
$E_{\rm H}$ is the ionization energy of the hydrogen ground state. A 
generalized version of this formula in employed for collisions between bands
of iron-group atoms (Werner \& Dreizler 1999). Forbidden CBB transitions are
calculated following Butler ({\tt http://ccp7.dur.ac.uk/Docs/detail.ps}):
\begin{equation}
C_{ij} = \frac{8.631\times 10^{-6}\, n_e\, e^{-u_0}\,\Omega}{g_i\,\sqrt{T}}
\end{equation}
where
\begin{equation}
\Omega \equiv \sum_{i=1}^{\sc nfit} a_i\,x^{i-1}
\end{equation}

\noindent
The fit parameters $a_{\rm 1}$, \ldots , $a_{\sc nfit}$ for the {\it effective
collision strength} are input parameters, $x=\log T - T_1$. 

The RBF cross-section are treated in various manners. For iron and calcium
detailed calculated data of the Opacity Project (The Opacity Project Team,
1995, 1997) are available. For the rest of the iron-group elements hydrogenic
cross-sections are used. For the CBF cross-sections no calculated data exist
for the iron group. Therefore they have to be estimated using the average RBF
cross-section and the Seaton (1962) formula.

The number of frequency points can be reduced by a factor of 1,000 by using
the statistical approach of the opacity sampling method. This method has been
described by Peytremann (1974) for LTE atmospheres. Anderson (1991) and
Dreizler \& Werner (1992) then used this idea for the calculation of iron-group
cross-sections. This Ansatz guarantees that the number of NLTE levels and
frequency points required for model-atmosphere calculations is reduced by
several orders of magnitude. For example a typical iron model atom
(Fe {\sc iv} -- {\sc vi}) is represented by  22 NLTE levels, which are
calculated from 16,000 individual levels, and 83 superlines which are
calculated from 2,200,000 individual lines.

For a direct comparison between observed and calculated spectra, however,
a subsequent calculation of a formal solution with a sufficiently fine
frequency grid is mandatory.

\section{Model Atmospheres and Synthetic Spectra}

The \textsc{pro2} NLTE model atmospheres are calculated in the most efficient
way with very detailed model atoms (Rauch 1997), consideration of HMF, and
Stark line broadening (Werner 1996; Rauch 2000). Moreover, it is sufficient to
treat linear Stark effect by approximate formulae (Uns\"old 1968; Werner, 
Heber, \& Hunger 1991)
\begin{equation}
\kappa(\Delta\lambda) = \frac{\pi e^2}{m c^2}\lambda^2 f \frac{1}{s^*_n F_0}
U\left(\frac{\Delta\lambda}{s^*_n F_0}\right)   
\end{equation}
with the microfield 
\begin{equation}
F_0 = 2.61 e\,\left[\,\sum_{\rm ions }^{~}z_i^{3/2}n_i\,\right]^{2/3}.
\end{equation}
$U(\beta)$ is given by van Dien (1949).
$s_n=0.0192\lambda^2[\,n_{\rm up}(n_{\rm up}-1)+n_{\rm low}(n_{\rm low}-1)\,]/Z$
is a measure of the width of the Stark pattern.

However, for the subsequent calculation of synthetic spectra even more detailed
model atoms (including fine-structure splitting) and a finer frequency grid are
necessary. Stark-broadening tables calculated for H\,I by Lemke (1997) and 
for He\,II by Sch\"oning \& Butler (1989), based upon VCS theory (Vidal, 
Cooper, \& Smith 1973) need to be used. For some metal ions, broadening tables
are published e.g.~by Dimitrijevi\'c, Sahal-Br\'echot, and others; check e.g.
\linebreak{\tt http://adsabs.harvard.edu/abstract\_service.html} for many references. 
\linebreak For fine-structure splitting, we use the energies given by Bashkin \& Stoner
(1975, 1978). For the calculation of oscillator strengths for the respective
transitions, we follow Kuhn (1969).

\section{A Database of Ready-To-Use Model Atoms}

When we started to work on this talk, we thought about the total time that we
have spent since 1986 in the search, collection, and evaluation of atomic data,
followed by the construction of model atoms and their extensive testing in
model-atmosphere calculations. Because the observational techniques have
improved tremendously in this time frame, and satellites such as FUSE and
XMM-NEWTON provide spectra of superb quality in the high-energy range, the work
on our model atoms continues steadily. We estimate that at least five man-years
have been used for this purpose --- in our group alone. 
Similar work has surely been done in all other model-atmosphere groups.
Unfortunately, there is no standard format yet for the model atoms. This fact
makes it difficult to compare results of model-atmosphere calculations from
two different codes because the atomic data-file cannot be exchanged. Moreover,
it appears to be both an enormous waste of man-power and a source of
uncertainty. An attempt to concentrate on the construction of model atoms in a
standard format, and collect these in a ``database'' which will be accessible
via WWW {\tt (http://astro.uni-tuebingen.de/$^\sim$rauch/ModelAtoms\_ready2use.html)}\linebreak
would be major progress in this field. Consequently, we propose such a database,
henceforth called the T\"ubingen Model-Atom Database ({\sc tmad}), with a
standard ASCII format which is adopted from our \textsc{pro2} format
(\S 4.1).
Creators of model atoms in the format required by the database may add their
files. They must be accompanied by a description, which indicates the relevant
parameter range (effective temperature, surface gravity), and references, where
the model atoms have been used in spectral analyses.

\subsection{Description of the \textsc{pro2} Atomic Data Files}

\textsc{pro2} expects an atomic-data file, which contains model atoms with
atomic data, or file names of atomic data. It is first created by the user and
then processed by an auxiliary program which partly checks the file for
consistency. A few keywords (Table 1) are used to indicate different sections
of the atomic data file (every section is closed by a single ``{\tt 0}'' in the
first column of a line). 

\begin{table}[ht]
\caption{Keywords in a \textsc{pro2} atomic data file}\vspace{2mm}
\begin{tabular}{lll}
Keyword &\multicolumn{2}{l}{Meaning} \\
\hline		
\noalign{\smallskip}
{\tt ATOM}&\multicolumn{2}{p{11cm}}{Introduces a new element. Its ions
in increasing order follow.\newline
The subsequent line indicates the chemical abbreviation ({\tt FORMAT A2}),
\newline the charge of the lowest ionization stage (in $e^-$) in the model atom,\newline
and the atomic weight (in AMU).} \\
\hline
\noalign{\smallskip}
{\tt L} & \multicolumn{2}{l}{NLTE levels} \\
{\tt LTE} & \multicolumn{2}{l}{LTE levels} \\
\hline
\noalign{\smallskip}
{\tt RBB} & radiative & bound-bound transitions \\
{\tt RBF} & radiative & bound-free transitions\\
{\tt RDI} & radiative & di-electronic transitions\\
{\tt RFF} & radiative & free-free transitions\\
{\tt RLL} & ``sample''  & bound-bound (within one band) \\
{\tt RLU} & ``sample''  & bound-bound (between two bands) \\
{\tt CBB} & collisional & bound-bound transitions\\
{\tt CBF} & collisional & bound-free transitions\\
{\tt CBX} & collisional & bound-bound transitions (NLTE to LTE levels) \\
\hline
\end{tabular}
\end{table}
\begin{table}[ht]
\caption{Example of a hydrogen model atom. Note that level \quad\quad\quad
energies are given in frequency units and cross-sections in cm$^2$.}\vspace{2mm}
{\tt ATOM\vspace{-0.4mm}\\
H\ \ 0\ \ 1.008\\
L\vspace{-0.4mm}\\
H11\ \ \ \ \ \ \ H21\ \ \ \ \ \ \ \ 3.2880912929E+15     2\vspace{-0.4mm}\\
H12\ \ \ \ \ \ \ H21\ \ \ \ \ \ \ \ 8.2202193884E+14     8\vspace{-0.4mm}\\
0\\
LTE\vspace{-0.4mm}\\
H13\ \ \ \ \ \ \ H21\ \ \ \ \ \ \ \ 3.6534150835E+14    18\vspace{-0.4mm}\\
0\\
RBB\vspace{-0.4mm}\\
H11\ \ \ \ \ \ \ H12\ \ \ \ \ \ \ \ 1 1   4.1620E-01\vspace{-0.4mm}\\
0\\
RBF\vspace{-0.4mm}\\
H11\ \ \ \ \ \ \ H21\ \ \ \ \ \ \ \ 1 3  7.91857E-18  1.  3.\vspace{-0.4mm}\\
H12\ \ \ \ \ \ \ H21\ \ \ \ \ \ \ \ 1 3  1.58372E-17  1.  3.\vspace{-0.4mm}\\
0\\
RFF\vspace{-0.4mm}\\
H1\ \ \ \ \ \ \  \ \ \ \ \ \ \ \ \ \ \ \  2 0\\
CBB\vspace{-0.4mm}\\
H11\ \ \ \ \ \ \ H21\ \ \ \ \ \ \ \ 1 0\vspace{-0.4mm}\\
0\\
CBF\vspace{-0.4mm}\\
H11\ \ \ \ \ \ \ H21\ \ \ \ \ \ \ \ 1 0\vspace{-0.4mm}\\
H12\ \ \ \ \ \ \ H21\ \ \ \ \ \ \ \ 1 0\vspace{-0.4mm}\\
0\\
CBX\vspace{-0.4mm}\\
H11\ \ \ \ \ \ \ H13\ \ \ \ \ \ \ \ 3 0\vspace{-0.4mm}\\
H12\ \ \ \ \ \ \ H13\ \ \ \ \ \ \ \ 3 0\vspace{-0.4mm}\\
0\\
L\vspace{-0.4mm}\\
H21\ \ \ \ \ \ \ NONE\ \ \ \ \ \ \ 0.0000000000E+00     1\vspace{-0.4mm}\\
0\\  
}
\end{table}
\textsc{pro2} level names are encoded in a 10-character string. This is a
reminder of the early 80's when we had to punch FORTRAN cards and wanted to
keep all information about one level on one card, i.e.~we could use only 66
characters. For ``simple'' elements this number was enough; but in the meantime
we have realized that a 40-character string for the level-names would be much
more convenient, for example:\vspace{2mm}\\
\noindent
{\tt MG07\uu 1s22s2p2\uu\uu\uu\uu\uu\uu\uu\uu\uu\uu\uu\uu 2DE\uu 3s\uu 1DO\uu\uu 3/2} 
\vspace{2mm}\\
allows one to identify unambiguously the element, ion, core- and 
valence-electron configuration, parity, and quantum numbers. Presently the
rudimentary designation {\tt MG73S'\uu 1D\uu} is used.
\vspace{1mm}\\
A detailed description of this data format and of the formulae which are used
for the transitions is given in the \textsc{pro2} User's Guide
(Werner, Rauch, \& Dreizler 2001). For all levels, the form for a general entry
is:\vspace{1mm}\\
{\tt Level\ \ \ \ \ Parent\ \ \ \ $E$\ $G$}
\vspace{1mm}\\
with the names of the level and its parent level (ground state of the next
ionization stage), the ionization energy $E$ (given in Hz), and the statistical
weight $G$ of the level.
For all transitions, the form for a general entry is\vspace{2mm}\\
\noindent
{\tt
Level$_{\tt{l}}$\ \ \ \ Level$_{\tt{u}}$\ \ \ \ $n_{\tt{form}}$\ $n_{\tt{dat}}$ P$_1$ $\ldots$ P$_{\tt n_{dat}}$
}
\vspace{2mm}\\
with the names of the lower (${\rm Level}_{\,\tt{l}}$) and upper
(${\rm Level}_{\,\tt{u}}$) level, the number $n_{\,\tt{form}}$ of the formula
which is used to calculate the cross-section, the number $n_{\,\tt{dat}}$ of
input parameters that follow, and the input parameters, respectively. In the
case of ``sample'' cross-sections, these are read in from data files which have
been produced by {\sc IrOnIc} (\S 4).
A very simple model atom for hydrogen which can be used to calculate a
(not very realistic) model atmosphere is shown in Table 2.

\section{Conclusions}

Present NLTE model-atmosphere codes have reached a very high level of
sophistication.  Now, strong efforts to achieve adequate atomic data must be
continued in order to be able to analyze reliably the high-quality spectra
which are already available from the infrared to the X-ray range. {\it Everyone
who is calculating stellar atmospheres should be aware of the important role
that atomic data plays, because one thing is for sure -- even if you use a
perfect code, if you put rubbish in, you will get rubbish out.}

\acknowledgments  
This research was supported by the DLR under grant 50\,OR\,0201 (TR)
and by the DFG under grant We1312/23-1 (JLD).

\end{document}